\documentstyle[11pt,psfig,newpasp,twoside]{article}

\begin{document}

\title{Revised Baade-Wesselink Analysis of RR Lyrae Stars}  

\author{Carla Cacciari$^1$, Gisella Clementini$^1$, Fiorella Castelli$^2$ 
and Fabrizio Melandri$^3$}
\affil{$^1$ Osservatorio Astronomico, Via Ranzani 1, I-40127 Bologna, Italy}
\affil{$^2$ Osservatorio Astronomico, Via Tiepolo 11, I-34131 Trieste, Italy}
\affil{$^3$ Dipartimento di Astronomia, Via Ranzani 1, I-40127 Bologna, Italy}

\begin{abstract}
We have applied the Baade-Wesselink method to two field RR Lyrae stars, 
i.e. SW And and RR Cet, and 
derived their distances and physical parameters. With respect to previous 
B-W analyses we have applied the following improvements: 
a) use of all sets of available data, after proper comparison for homogeneity 
and compatibility; b) use of the most recent and accurate model atmospheres,   
with turbulent velocity V$_{turb}$ = 4 km/s and the no-overshooting 
approximation, and comparison with other treatments of 
convection; c) use of the instantaneous gravity along the pulsation cycle 
rather 
than the mean value; d) comparison with modified radial velocity curves 
according to various assumptions on radial velocity gradients in the atmosphere; 
and e) careful reanalysis of the temperature scale.
The main aim of this study is to evaluate the effect of the above items  
on the B-W results and verify whether any (or a combination) of them 
can possibly account for the discrepancy of the absolute magnitude zero-point
with respect to other independent determinations.   
\end{abstract}


\section{Introduction}

The distance scale problem is not 
solved yet, the dichotomy between {\it long} and {\it short} distance scales 
becoming even more clear-cut. 
In favour of the {\it short} distance scale (i.e. faint Mv(RR),  $(m-M)_{LMC} \le 18.3$) 
are the studies on RR Lyrae statistical parallaxes (Gould \& Popowski 1998), 
local RR Lyrae kinematics (Martin \& Morrison 1998), 
eclipsing binaries in the LMC (Udalski et al. 1998, but see also Guinan et al. 1998 for a 
contrasting result), 
red clump stars (Cole 1998), Hipparcos parallaxes of field HB stars (Gratton 1998). 
On the other hand, in favour of the {\it long} distance scale (i.e. bright Mv(RR), 
$(m-M)_{LMC} \ge 18.5$) are the studies on  sd-MS fitting in globular clusters 
(Gratton et al. 1997; Reid 1997; Chaboyer et al. 1998; Pont et al. 1998; 
Grundahl, VandenBerg, \& Andersen  1998; Carretta et al. 1999), 
RGB bump stars in globular clusters (Ferraro 
et al. 1999), tip of the RGB (Lee, Freedman \& Madore 1993), 
RR Lyraes period-shift effect (Sandage 1993), RR Lyraes double-mode pulsators 
(Kovacs \& Walker 1998), Hipparcos parallaxes of Cepheids (Feast \& Catchpole 1997), 
SN1987A in the LMC (Panagia 1998).

The Baade-Wesselink (B-W) analysis of RR Lyrae stars can help solve this problem. 
For a review and general description of this method and its assumptions and approximations 
see for example Gautschy (1987). 

Previous B-W analyses were performed on 29 field RR Lyraes by a few independent groups, 
e.g. Liu \& Janes (1990), Jones et al. (1992), Cacciari, Clementini \& Fernley (1992), 
and Skillen et al. (1993).  
These references are only indicative and are by no means exhaustive or 
complete. 
The results of these and several more B-W studies have been reanalysed and summarised 
by Fernley (1994) and Fernley et al. (1998) who find a relation 
$M_V(RR) = 0.20 \pm 0.04 [Fe/H] + 0.98 \pm 0.05$ between the absolute visual magnitude 
and the metallicity of the RR Lyrae stars. This result is intermediate but closer to the 
short distance scale, corresponding to $(m-M)_{LMC} \sim 18.34$.  

The previous B-W analyses were based on: 
i) Kurucz (1979) model atmospheres, with turbulent velocity $V_{turb}$ = 2 km/s; 
ii) semi-empirical temperature calibration; 
iii) mostly IR (i.e. K), but also visual V,R and I photometry; 
iv)  the use of an average value for gravity (usually log$g$=2.75); 
v)   a restricted phase range for fitting (usually 0.35 to 0.80); 
vi)  the barycentric (also called $\gamma$-) velocity derived as simple integration 
over the entire pulsation cycle; 
vii) a constant factor (1.38) to transform radial into pulsational velocities. 

Several model atmospheres are now available, with:
improved {\it opacities}; different treatments of {\it convection}; 
different values of {\it turbulent velocity}; more accurate {\it temperature and 
BC calibrations}. 
The aim of the present work is to test what is the effect of these new models and 
calibrations, as well as of other assumptions, on the B-W results for RR Lyrae stars. 

\section{Our re-analysis: assumptions and approximations}

\subsection{New elements}  

{$\bullet$ \bf Model atmospheres.} We have used the following sets of model atmospheres: \\
i) Kurucz (1995) with MLT+overshooting treatment of convection, $V_{turb}$ = 2 and 4 km/s, 
[m/H]=0.0; 
ii) Castelli (1999b) with MLT without overshooting treatment of convection, [m/H]=0.0 and 
$V_{turb}$ = 2 km/s, and [m/H]=--1.5 and $V_{turb}$ = 2 and 4 km/s. The models at [m/H]=--1.5
are enhanced in $\alpha$-elements by [$\alpha$/$\alpha_{\sun}$]=+0.4. 
iii) Castelli (1999a)  experimental models  with no convection. These models do not have 
physical meaning, but are only intended to mimick the effects of 
recent treatments of convection, e.g. MLT with l/H=0.5 instead of 1.25 (Fuhrmann, Axer,
\& Gehren 1993), 
or Canuto \& Mazzitelli (1992) approximation that predicts a very low convection or zero 
convection for stars with $T_{eff} \ge$ 7000 K and that has 
been suggested to provide a better match to the data (see Gardiner, Kupka, 
\& Smalley 1999 for a recent re-discussion of this issue). 
The no-convection models are available for [m/H]=0.0 and --1.5 
and $V_{turb}$ = 2 km/s. 

\noindent
{$\bullet$ \bf Gravities.} The values of log$g$ have been calculated at each phase-step 
from radius percentage variation (assuming $\Delta$R/$<R> \sim$ 15\%) plus the acceleration 
component derived from the radial velocity curve. The zero-point was set from theoretical 
ZAHB models, i.e. log$g$=2.86 at the phase corresponding to average radius. 
Note that all ZAHB models give average log$g$=2.86 $\pm$ 0.01 (Dorman, Rood, \& O'Connell 
1993; Sweigart 1997 with no He-mixing; Chieffi, Limongi \& Straniero 1998) with the 
only exception of Sweigart (1997) models with He-mix=0.10 which give log$g$=2.75 but may 
not be applicable to our field stars. 

\noindent
{$\bullet$ \bf Semi-empirical Teff and BC calibration.} We have used Montegriffo et al. 
(1998) semi-empirical $BC_V$ and $BC_K$ calibration 
and temperature scale for PopII giants, that  is based on RGB and HB stars in 10 globular 
clusters. 

\noindent
{$\bullet$ \bf Gamma-velocity.}  The default value of the  
$\gamma$-velocity was estimated from integration of the observed radial velocity curve over 
the entire pulsation cycle, as it was done in all previous B-W studies. However, 	   
Oke, Giver, \& Searle (1962) had suggested for SU Dra that velocity gradients may 
exist in the atmosphere, and proposed to take them into account by  correcting the observed 
radial velocities  of a 
positive quantity that would vary about linearly between phase 0.95 and 0.40. 
On the other hand, Jones et al. (1987)  found no observational  evidence of 
velocity gradients among weak metal lines in X Ari, at least within their observational 
errors ($\pm$ 2 km/s). Chadid \& Gillet (1998), however, did seem to find some evidence of 
differential velocities among weak metal lines in RR Lyr:  the radial velocity curve from the 
FeII($\lambda$4923.9) line shows a slightly larger amplitude than that from the 
FeI($\lambda$4920.5) line which forms a little deeper in the atmosphere. A similar effect 
was found between BaII and TiII lines, and is related to the presence of strong shocks. \\
Therefore, 
in addition to the default $\gamma$-velocity calculation from the observed RV curve, 
we have simulated two other cases,  $\gamma$-1 where the RV 
curve has been corrected as suggested by Oke et al. (1962) albeit by a much smaller amount 
(+ 2.0 km/s at most), and $\gamma$-2 where the amplitude of the RV curve has been 
stretched by $\pm$ 5 km/s at the phases of maximum and minimum RV. 
These simulations are only numerical experiments and do not intend to provide realistic  
answers to the problem of radial velocity gradients in the atmosphere: the radial velocities 
for the RR Lyrae stars are derived from a large number of weak metal lines and it is still 
totally unclear which correction (if any) should be applied to the average values in the 
presence of velocity gradients among some of these lines. 

\subsection{Old Assumptions}

{$\bullet$ \bf The $p$-factor.} The factor used to transform radial into pulsational velocity
has been assumed to be 1.38, as in Fernley (1994). It might be a few \% smaller, this 
conservative assumption gives the brightest possible luminosity as a function of $p$. 
 
\noindent	  
{$\bullet$ \bf Fitting phase interval.} As with all previous applications of the B-W 
method, the fitting has not been performed  on the entire 
pulsation cycle, but on a restricted phase interval appropriate for each star:
0.30-0.80 for SW And, for best stability of results; and 0.25-0.70 for RR Cet, to avoid 
shock-perturbed phases. 

\section{Results on SW And and RR Cet, and Discussion} 
\subsection{SW And}

We have used all possible and compatible data from the literature, i.e. BVRIK and uvby 
photometry, and radial velocities.
The adopted input parameters for this star are [Fe/H]=0.0, E(B--V)=0.09, 
$\gamma$-velocity=--19.94 km/s and $<V_0>$=9.44.

The default model calibration on Vega led to systematically hotter 
temperatures (by $\sim$ 180K) from the (B--V) colors with respect to all other 
colors. This difference has no physical justification, and is only due to a 
behaviour of the models that was already known and commented by Castelli (1999b). 
It is worth mentioning that this hotter temperature by $\sim$ 180K leads to a brighter 
$M_V$ magnitude by $\sim$0.12 mag using the combination V and B$-$V in the B-W analysis, 
and by only $\sim$0.02 mag using K and B$-$V. 
{\it It is important to note that the use of the K magnitude with any color, in 
particular V$-$K, is the least affected by temperature uncertainties, and provides 
the most stable results.}

In Table 1 we summarize the values of $M_{V}$ that result from the use of the K 
magnitude and the various colors and models. The models are labelled as: K2 and K4 
for Kurucz (1995) with $V_{turb}$= 2 and 4 km/s respectively; K2-nover and K2-noconv 
for Castelli (1999a,b) $V_{turb}$= 2 km/s and the no-overshooting and no-convection 
approximations respectively; Montegriffo for the Montegriffo et al. (1998) temperature 
scale and $BC_K$ calibration. 

\begin{table}[htb]
\caption{Determinations of $M_{V}$ for SW And}
\begin{center}
\begin{tabular}{lcccccc}
\tableline
Model       & ~~b$-$y~~ & ~~B$-$V~~ & ~~V$-$R~~ & ~~V$-$I~~ & {\bf ~~V$-$K~~} & ~~R$-$K~~ \\
\tableline
K2          & ~~1.11~~ & ~~0.83~~ & ~~1.02~~ &~~1.07~~&{\bf ~~0.93~~}& ~~0.97~~ \\
K4          & ~~1.09~~ & ~~0.91~~ & ~~1.03~~ &~~1.05~~&{\bf ~~0.92~~}& ~~0.95~~ \\
K2-nover    & ~~1.08~~ & ~~0.84~~ & ~~0.93~~ &~~0.99~~&{\bf ~~0.96~~}& ~~0.97~~ \\
K2-noconv   & ~~0.62~~ & ~~0.40~~ & ~~0.46~~ &~~0.51~~&{\bf ~~0.48~~}& ~~0.49~~ \\
Montegriffo &            &            &            &          &{\bf ~~0.94~~}&            \\
\tableline
\tableline
\end{tabular}
\end{center}
\end{table}

We note that: 

\noindent
$\bullet$ The use of the b$-$y colors yields unreliable results because the angular diameter 
curve is distorted and the fitting to the linear radius curve is poor.

\noindent
$\bullet$ The no-convection models yield much brighter magnitudes, 
the exact amount of ``brightening'' depending on the details of the convection treatment. 

\noindent
$\bullet$ All models, except the no-convection ones, yield consistent results with the 
empirical relation by Montegriffo et al. (1998). 

\noindent
$\bullet$ The present results, except the no-convection ones, are consistent with the 
previous determination of $M_V$ for SW And, i.e. {\bf 0.94} (Fernley 1994). 

\subsection{RR Cet}

We have used all compatible BVRIK photometry and radial velocities from the literature. 
The adopted input parameters for this star are [Fe/H]=--1.5, E(B--V)=0.05, 
$\gamma$-velocity=--74.46 km/s and $<V_0>$=9.59.

The default model calibration on Vega for the [Fe/H]=--1.5 models 
led to consistent temperatures within $\sim$ 100K from all colors. 
 
In Table 2 we summarize the values of $M_{V}$ for RR Cet that result from the use of the K 
magnitude and the various colors and models. In addition to the models described also in 
Table 1, we have here a few other cases: K4-nover for Castelli (1999b) $V_{turb}$= 4 km/s 
and the no-overshooting approximation; K2/K4-nover for $V_{turb}$ = 4 km/s only at phases 
0.60-1.00 and $V_{turb}$ = 2 km/s elsewhere; $\gamma$-1 for a corrected RV curve 
by + 2 km/s at the phase of minimum RV; $\gamma$-2 for a corrected RV curve 
whose amplitude has been stretched by $\pm$ 5 km/s.

\begin{table}[htb]
\caption{Determinations of $M_{V}$ for RR Cet}
\begin{center}
\begin{tabular}{lccccc}
\tableline
Model       & ~~B$-$V~~ & ~~V$-$R~~ & ~~V$-$I~~ & {\bf ~~V$-$K~~} & ~~R$-$K~~ \\
\tableline
K2-nover    & ~~0.54~~ & ~~0.44~~ & ~~0.47~~ & {\bf ~~0.57~~} & ~~0.60~~ \\
K2-noconv   & ~~0.53~~ & ~~0.42~~ & ~~0.45~~ & {\bf ~~0.55~~} & ~~0.59~~ \\
K4-nover    & ~~0.51~~ & ~~0.41~~ & ~~0.45~~ & {\bf ~~0.55~~} & ~~0.59~~ \\
K2/K4-nover & ~~0.64~~ & ~~0.48~~ & ~~0.51~~ & {\bf ~~0.58~~} & ~~0.60~~ \\
K4-nover $\gamma$-1   & ~~0.54~~ & ~~0.44~~ & ~~0.48~~ & {\bf ~~0.58~~} & ~~0.61~~ \\
K4-nover $\gamma$-2   & ~~0.35~~ & ~~0.25~~ & ~~0.30~~ & {\bf ~~0.39~~} & ~~0.43~~ \\
Montegriffo           & ~~0.48~~ &            &            & {\bf ~~0.56~~} &      \\
\tableline\tableline
\end{tabular}
\end{center}
\end{table}

We note that: 

\noindent
$\bullet$ The no-convection models seem to have no significant effect on the derived 
magnitudes. 

\noindent
$\bullet$ The use of $V_{turb}$ = 4 km/s all over the fitting phase interval  
yields insignificantly brighter magnitudes (by $\le$ 0.03 mag) with respect to the case 
at $V_{turb}$ = 2 km/s. On the other hand, setting $V_{turb}$ = 4 km/s only at phases 
0.60-1.00 and $V_{turb}$ = 2 km/s elsewhere leads to somewhat fainter magnitudes, 
the effect being stronger in the bluer colors. 

\noindent
$\bullet$ Correcting the RV curve as suggested by Oke et al. (1962) by at most + 2 km/s 
at the phase of minimum RV (case $\gamma$-1) leads to slightly fainter magnitudes. 

\noindent
$\bullet$ Correcting the RV curve by stretching the amplitude in analogy to the behaviour of 
FeII and BaII lines (case $\gamma$-2) leads to brighter magnitudes. The amount 
of brightening we have estimated is however too large, since the correction we have 
applied ($\pm$ 5 km/s) was exaggerated for the sake of computation. 

\noindent
$\bullet$ All models yield consistent results with the empirical relation by Montegriffo 
et al. (1998). 

\noindent
$\bullet$ The present results  are slightly brighter than the 
previous determination of $M_V$ for RR Cet, i.e. {\bf 0.68} (Fernley 1994). 

\subsection{Conclusions} 

\noindent
$\bullet$ 
The meaning of our simulations  with the no-convection models is only to point out that, 
if the effect of convection were indeed overestimated by the  present MLT, 
a more correct treatment with a reduced  
impact of convection would lead to brighter magnitudes for solar-metallicity stars, but would 
have lesser consequences on metal-poor stars. However, Gardiner et al. (1999) reach no 
definitive conclusion on this subject, and suggest that the classical treatment of convection 
with l/H=1.25 may still be the best approach in the temperature range 6000-7000K which is 
relevant for RR Lyrae stars around minimum light. 

\noindent
$\bullet$ The corrections to the radial velocity curves simulated with RR Cet are quite 
arbitrary and not sufficiently supported by observational evidence. They are only intended 
as numerical experiments to test their effect on the derived magnitudes. 

\noindent
$\bullet$ The present results are quite compatible with the results from previous analyses, 
and do not support a significantly brighter zero-point for the RR Lyraes luminosity scale.

\end{document}